\documentclass[12pt]{iopart}
\usepackage{amsfonts}
\usepackage{iopams}

\newtheorem{theorem}{Theorem}
\newtheorem{corollary}[theorem]{Corollary}
\newtheorem{definition}[theorem]{Definition}
\newtheorem{proposition}[theorem]{Proposition}
\newtheorem{remark}[theorem]{Remark}
\newenvironment{proof}[1][Proof]{\textbf{#1.} }{\ \rule{0.5em}{0.5em}}

\newcommand{\text}[1]{{\mathrm{#1}}}
\newcommand{\HH}{{\mathcal{H}}}
\newcommand{\BB}{{\mathcal{B}}}
\newcommand{\LL}{{\mathcal{L}}}

\newcommand{\real}{{\mathbb R}}

\newcommand{\ZZ}{{\widetilde{Z}}}
\newcommand{\EE}{{\widetilde{E}}}

\begin{document}

\title[Phase space measurements]{A complete characterization of phase space measurements}

\author{C~Carmeli\dag, G~Cassinelli\dag, 
E~DeVito\ddag, A~Toigo\dag\  and B~Vacchini\S}

\address{\dag\ Dipartimento di Fisica, Universit\`a di Genova, and I.N.F.N.,
Sezione di Genova, Via Dodecaneso~33, 16146 Genova, Italy}
\address{\ddag\ Dipartimento di Matematica, Universit\`a di Modena e Reggio
Emilia, Via Campi 213/B, 41100 Modena, Italy, and I.N.F.N., Sezione di
Genova, Via Dodecaneso~33, 16146 Genova, Italy}
\address{\S\ Dipartimento di Fisica, Universit\`a di Milano, and I.N.F.N.,
Sezione di Milano, 20131, Italy}

\eads{{\mailto{carmeli@ge.infn.it}}, {\mailto{cassinelli@ge.infn.it}}, {\mailto{devito@unimo.it}}, 
{\mailto{toigo@ge.infn.it}}, {\mailto{vacchini@mi.infn.it}}}

\begin{abstract}
We characterize all the phase space measurements for a non-relativistic particle.
\end{abstract}

\submitto{\JPA}

\section{Introduction}
In the usual framework of Quantum Mechanics, the states (density
matrices) of a physical system are described by positive trace
class trace one operators acting on a Hilbert space $\HH$, and
the physical quantities (observables) are associated with
self-adjoint operators on $\HH$ in such a way that ${\rm tr}(SA)$
is the expectation value of the observable $A$ when the system is
in the state $S$ (here ${\rm tr}$ denotes the trace).

Nevertheless, a careful analysis of measurement processes shows that
one has to generalize suitably the concept of observable for both
theoretical and experimental reasons~\cite{Ludwig-Foundations,
  Hol,pippo2,HolevoNEW}. 
These {\em generalized observables} are
described as mathematical objects by {\em positive operator valued
  measures} (POVM). 
In this  framework one can describe
measurements of quantities like angle of rotation, phase and arrival
times, as well as joint measurements of quantities like position and
momentum, incompatible according to the standard textbook formulation
of Quantum Mechanics.

In order to give a physical meaning to the observables one invokes some properties of covariance 
with respect to a symmetry group. 
The requirement of covariance is a
strong constraint: it allows to select the measurements of physical
interest among the larger class of all the possible generalized
observables. 
As recently proved, from this principle it follows not only the characterization of generalized 
observables, but also  the determination of generators of quantum dynamical
semigroups~\cite{HolevoJMP,art,art1}.

In this paper, we classify all the possible joint observables of
position and momentum that arise from the request of covariance
with respect to the Galilei group. In literature these observables
are usually called {\em phase space measurements} for a
non-relativistic  particle. We restrict our attention to the isochronous Galilei 
group since the POVMs covariant with respect to this group have a clear and 
transparent physical meaning. Moreover the technical treatment (compare Prop.~4) 
is rather simple. We have in mind the possibility of treating more general 
space-time groups (e.g. Poincar\'e, de~Sitter)

The quest for the characterization of phase space measurements
in Quantum Mechanics goes back to the 70's, and in particular to the
seminal works of Ali and Prugove\v{c}ki~\cite{Ali-xp} and
Holevo~\cite{Holevo-xp,Hol}, the first concerned with the
representation of Quantum Mechanics on fuzzy phase space, the second
with a general treatment of quantum measurements covariant with
respect to a given symmetry group. The result presented in this paper,
which relies on a previous work on the characterization of POVM
covariant with respect to an irreducible representation of a symmetry
group~\cite{Nostro}, essentially confirms the previous ones showing,
along a different line of proof, that indeed all phase space
  measurements for a non-relativistic  particle are expressed
in terms of an operator valued density, thus releasing the more
restrictive assumptions considered in~\cite{Ali-xp} (see
also~\cite{pippo2}) and putting into evidence with respect
to~\cite{Holevo-xp,Hol} that square-integrability of the considered
representation is both a sufficient and necessary condition. 

The paper is organized in the following way. In Section
\ref{storia} we briefly review the physical motivations that
justify the introduction of covariant positive operator valued
measures from the point of view of quantum measurement theory. In
Section \ref{main} we give the complete classification of the
phase space measurements for a non-relativistic particle.
The proof of the result is given in Section \ref{proof}.

\section{A brief review on POVMs}\label{storia}

For an exhaustive exposition of the theory of covariant POVMs from
the perspective of quantum measurement theory, one can refer to
\cite{Hol,pippo2,HolevoNEW}. However, for the reader's convenience, we briefly
recall the main steps which lead quite naturally to the idea of
covariant POVM.

First of all, we recall the mathematical definition of POVM.
\begin{definition}\label{povm}
Let $X$ be a metric space and $\HH$ a (complex separable) Hilbert space. A map $E$ from the Borel 
subsets $\BB(X)$
of $X$ into the set $\LL(\HH)$ of bounded operators on $\HH$ such that:
\begin{enumerate}
\item $\langle \phi,E(Z)\phi\rangle\geq 0 \quad \forall\ \phi\in\HH,\,Z \in \BB(X)$
\item $E(X)=I $
\item $E(\cup_i Z_i)= \sum_i E(Z_i)$ for all disjoint sequences of subsets (the series converging 
in the weak sense).
\end{enumerate}
is called a (normalized) positive operator valued measure (POVM) based on $X$.
\end{definition}

The role of POVMs in Quantum Mechanics is justified by the
following observation. Given a physical
quantity described by a self-adjoint operator $A$, it is well known how one obtains
the probability distribution of the outcomes of $A$. Indeed  by
the spectral theorem, $A$ uniquely defines a {\em
projection valued measure} $P$, i.~e.~a map 
\begin{equation}
P:\, \BB(\real)\rightarrow \LL (\HH)
\end{equation}
from the Borel subsets $\BB(\real)$ of $\real$ into the space
of bounded operators  $\LL (\HH)$ on $\HH$ satisfying the following
three conditions:
\begin{enumerate}
\item $P(Z)$ is an orthogonal projection operator for all $Z\in\BB(\real)$:
\begin{eqnarray}
\label{proiettore}
P(Z)=P^*(Z)=P(Z)^2 \quad \forall Z \in \BB(\real)
\end{eqnarray}
\item $P(\real)=I$
\item $P(\cup_i Z_i)= \sum_i P(Z_i)$ for all disjoint sequences of subsets (the series converging in the weak sense).
\end{enumerate}
Comparing with definition \ref{povm}, one easily checks that a
projection valued measure is a particular case of POVM.
With this notation, the physical content of Quantum Theory is
based on the following assumption: if one measures the observable
$A$ when the system is in a state $S$, the probability to have an
outcome in $Z$ is given by ${\rm tr}[SP(Z)]$.

The fact that $P$ is a projection valued measure assures
that the map
\begin{equation}
\label{trace} Z\mapsto  {\rm tr}[SP(Z)]=:\mu^A_S(Z)
\end{equation}
is a probability distribution on $\real$. Clearly,  the physical content of
the observable $A$ is completely given by the map
\[S\mapsto \mu^A_S\]
from the set of states into the space of probability measure on
$\real$ (the above map is usually called a {\em measurement}).

The key remark is that in order that equation~(\ref{trace}) defines a
probability measure, it is sufficient and necessary to replace equation~(\ref{proiettore})
with the weaker  condition that $P(Z)$ is a positive
operator, that is
\begin{eqnarray}
\label{positivo}
\langle \phi,P(Z)\phi\rangle\geq 0 \quad \forall\ \phi\in\HH,\, Z \in \BB(\real).
\end{eqnarray}
Then the corresponding map $Z\mapsto P(Z)$ will be a positive operator
valued measure on $\real$.

Moreover, in order to take into account joint measurements,
another generalization suggested by this approach consists in
assuming that the space of measurement outcomes is an arbitrary
metric space $X$ instead of $\real$. For example the joint
measurements of  position along the three axis of the Euclidean
space defines a projection measure on  $X= \real^3$.

Given a POVM $E$ on the space $X$, by the above discussion it is
reasonable to define a {\em generalized} measurement associated to
$E$ as a map 
from the set of states to the space of probability measures on
$X$
\[
   S\mapsto \mu_S^E,
\]
with $\mu_S^E$ defined according to equation~(\ref{trace}) 
\[Z\mapsto \tr[SE(Z)]=\mu_S^E(Z).\]

This mathematical framework can be further enriched introducing the
concept of POVM {\em covariant} with respect to a symmetry group. From
a mathematical point of view, one has the following definition.
\begin{definition}
Let $G$ be a group that acts both on $\HH$ by means of a projective unitary
representation $U$ and on the outcome space $X$ by a geometrical
(left) action $\alpha$. A POVM $E$ on $X$ is said to be {\em covariant} with
respect to $G$ if, for all $g\in G$,
\begin{equation}
\label{cov}
U_gE(Z)U_g^*\,=\,E(\alpha_g(Z)) \quad \forall Z\in \BB(X)\text{.}
\end{equation}
\end{definition}

In order to explain the physical meaning of equation~(\ref{cov}), let
us fix the ideas on a simple example and give a natural definition
of a position measurement on the real line $\real$, on which
$\real$ itself acts as the group of translations. If $x\in \real$,
its action on an element $y\in \real$ is $\alpha_x(y)=x+y$. If $S$
is a state, denote with $xS$ the translate of $S$ by $x$. In order
that a measurement $E$ be a position measurement, the probability
distribution of the outcomes of $E$ performed on $S$ and $xS$
should satisfy the following relation:
\begin{equation}
\mu^{E}_{xS}(Z+x)= \mu^E_S(Z) \quad \forall Z\in \BB(\real),\, x\in \real.
\end{equation}

In the more general setting in which a generic transformation
group $G$ acts both on the quantum system and on the outcome space
$X$ the above condition reads
\begin{equation}
\label{mu} \mu^{E}_{gS}(\alpha_g(Z))= \mu^E_S(Z)\quad \forall Z\in
\BB(X), g\in G.
\end{equation}
Since the action of $g\in G$ on the state $S$ is given by
\[
gS =U_g S U_g^*
\]
a straightforward calculation shows that equation~(\ref{trace}) and equation
(\ref{mu}) imply the covariance condition (\ref{cov}).

In particular, if $X$ is the (classical) phase space of the system
on which the isochronous Galilei group acts, the POVMs based on
$X$ and satisfying equation~(\ref{cov}) are called {\em phase space
measurements}.

\section{Phase space measurements}\label{main}

In the present section, we characterize all the phase space
measurements of a non-relativistic particle of mass $m$.
For the sake of simplicity we restrict to the spinless case, the extension to the general case 
being straightforward.

Every observer describes the phase space associated with a free
particle as $X=\mathbb{R} ^{3}\times \mathbb{P}^{3}$. The symmetry
group is the isochronous Galilei group $G=\left(
  \mathbb{R}^{3}\times \mathbb{V}^{3}\right) \times ^{\prime }SO\left(3\right) $, where 
$\mathbb{R}^{3}$ is the $3$-dimensional vector group of space
translations, $\mathbb{V}^{3}$ is the $3$-dimensional vector group of
Galileian boosts and $SO(3)$ is the group of rotations (connected with
the identity). In particular, the composition law of $G$ is given by
\begin{equation*}
\left( \vec{a},\vec{v},R\right) \left( \vec{
a}^{\prime },\vec{v}^{\prime },R^{\prime }\right) =\left(
\vec{a}+R\vec{a}^{\prime },\vec{v}+R
\vec{v}^{\prime },RR^{\prime }\right) \text{.}
\end{equation*}
The action of an element $g=\left(
  \vec{a},\vec{v},R\right) \in G$
on a point $\left( \vec{q},\vec{p}\right) \in X$
  is given by
\begin{equation}
\alpha_g \left( \vec{
q},\vec{p}\right) =\left( \vec{a}+R\vec{q}
,m\vec{v}+R\vec{p}\right) \text{.}  \label{azione di G}
\end{equation}

The Hilbert space  of a non-relativistic spinless particle of mass
$m$ is $\mathcal{H}=L^{2}\left(
\mathbb{R}^{3},\text{d}\vec{x}\right) $ and $G$ acts on
$\HH$ by means of the irreducible projective unitary
representation $U$ given by
\begin{equation}
\left[ U_{\left( \vec{a},\vec{v},R\right)} \phi
\right] \left( \vec{x}\right)
=e^{im\vec{v}\cdot \left(
\vec{x}-\vec{a}\right) }\phi \left(
R^{-1}\left( \vec{x}-\vec{a}\right) \right)
\text{.} \label{La rappr.}
\end{equation}
With these notations, the problem of determining the phase space measurements reduces
to the characterization of the POVM on $X$ covariant with respect to $U$.
The following theorem faces up this problem.
\begin{theorem}
\label{numero}
Let $T\in\BB(\HH)$ be a positive trace class trace one operator such that
\begin{eqnarray}
T U_{\left( \vec{0},\vec{0},R\right)} & = &
U_{\left( \vec{0},\vec{0},R\right)} T \ \
\forall R\in SO(3), \label{comm}
\end{eqnarray}
i.~e.~$T$ is a density matrix invariant under rotations.
For all $Z\in\BB(X)$ let $E_T(Z)$ be the operator
\begin{equation}
E_{T}\left( Z\right) =\frac{1}{\left( 2\pi \right) ^{3}}\int_{Z}U_{\left(
\vec{a},\frac{\vec{p}}{m},I\right)} T U_{\left( \vec{a},
\frac{\vec{p}}{m},I\right)} ^{\ast }\text{d}\vec{a}\text{d}
\vec{p}\text{.}  \label{L'altra}
\end{equation}
where the integral is understood in the weak sense.

The map $Z\mapsto E_T(Z)$ is a POVM on $X$ covariant with
respect to $U$.

Conversely, if $E$ is a POVM on $X$ covariant with respect to $U$,
then there exists a 
density matrix invariant under rotations such that $E=E_T$.
\end{theorem}

The proof of the above theorem (which is a special case of a more general
result~\cite{Nostro}) is given in the next section and it is based on the fact that
$G$ acts transitively on $X$, 
i.~e.~given any $x,y\in X$ it is always possible to find $g\in G$ such
that $\alpha_g(x)=y$. In particular, the stability subgroup at
the origin $(\vec{0},\vec{0})$, i.~e.~the
subgroup of elements of $G$ acting trivially on the origin, is the compact
group $SO(3)$, so that $X$ is isomorphic to the quotient space $G/SO(3)$.
The essential property involved in the proof of theorem~\ref{numero} is the fact that $U$ is  \em 
square-integrable\rm\ (see the definition in the next section). We will prove that 
square-integrability is a necessary and sufficient condition for the existence of covariant POVMs, when the stabilizer %%@
is compact.  

Equation~(\ref{L'altra}) can obviously also be written in terms of the Weyl operators according to
\begin{displaymath}
   E_{T}\left( Z\right) =\frac{1}{\left( 2\pi \right) ^{3}}\int_{Z}
e^{i (\vec{p}\cdot \vec{Q}-\vec{a}\cdot \vec{P})}
T
e^{-i (\vec{p}\cdot \vec{Q}-\vec{a}\cdot \vec{P})}
\text{d}\vec{a}\text{d}
\vec{p}\text{,}
\end{displaymath}
where $\vec{Q}$ and $\vec{P}$ denote position
and momentum operators acting in $L^{2}\left(
\mathbb{R}^{3},\text{d}\vec{x}\right)$.

We now characterize the positive trace class trace one operators
$T$ satisfying  equation~(\ref{comm}). We have the
factorization
\begin{equation*}
L^{2}\left( \mathbb{R}^{3},\text{d}\vec{x}\right) =L^{2}\left(
S^{2},\text{d}\Omega \right) \otimes L^{2}\left( \mathbb{R}_{+},r^{2}\text{d}r\right) \text{.}
\end{equation*}
Denoting with $l$ the representation of $SO\left(3\right) $ acting in $
L^{2}\left( S^{2},\text{d}\Omega \right) $ by left translations, we have
\begin{equation*}
\left. U\right| _{SO\left(3\right) }=l\otimes I\text{.}
\end{equation*}
The representation $\left( l,L^{2}\left( S^{2},\text{d}\Omega \right)
\right) $ decomposes into
\begin{equation*}
L^{2}\left( S^{2},\text{d}\Omega \right) =\bigoplus_{\ell \geq 0}M_{\ell },
\end{equation*}
where each irreducible inequivalent subspace $M_{\ell }$ is generated by the
spherical harmonics $\left( Y_{\ell m}\right) _{-\ell \leq m\leq \ell }$. We
have
\begin{equation*}
L^{2}\left( \mathbb{R}^{3},\text{d}\vec{x}\right) =\left(
\bigoplus_{\ell \geq 0}M_{\ell }\right) \otimes L^{2}\left( \mathbb{R}
_{+},r^{2}\text{d}r\right) =\bigoplus_{\ell \geq 0}\left( M_{\ell }\otimes
L^{2}\left( \mathbb{R}_{+},r^{2}\text{d}r\right) \right) .
\end{equation*}

Let $P_{\ell }:L^{2}\left( S^{2},\text{d}\Omega \right) \longrightarrow
L^{2}\left( S^{2},\text{d}\Omega \right) $ be the orthogonal projection onto
the subspace $M_{\ell }$. If $T$ intertwines $l\otimes I$, one has
\begin{equation*}
T\left( P_{\ell }\otimes I\right) =\left( P_{\ell }\otimes I\right) T\text{,}
\end{equation*}
where $P_{\ell }\otimes I$ projects onto $M_{\ell }\otimes L^{2}\left(
\mathbb{R}_{+},r^{2}\text{d}r\right) $. Given Hilbert spaces $\mathcal{H}_{1}
$ and $\mathcal{H}_{2}$ and an irreducible representation $\left( \pi , 
\mathcal{K}\right) $, a standard result asserts that $\mathcal{C}\left( \pi
\otimes I_{\mathcal{H}_{1}},\pi \otimes I_{\mathcal{H}_{2}}\right) =I_{ 
\mathcal{K}}\otimes \mathcal{L}\left( \mathcal{H}_{1},\mathcal{H}_{2}\right)
$. Since $M_{\ell }$ is irreducible, this implies
\begin{equation*}
T\left( P_{\ell }\otimes I\right) =P_{\ell }\otimes T_{\ell }
\end{equation*}
with $T_{\ell }\in \mathcal{L}\left( L^{2}\left( \mathbb{R}_{+},r^{2}\text{d} 
r\right) \right) $. We then have
\begin{equation*}
T=\sum_{\ell }T\left( P_{\ell }\otimes I\right) =\sum_{\ell }P_{\ell
}\otimes T_{\ell }\text{.}
\end{equation*}
In the last expression, $T$ is a positive trace one operator if and only if
each $T_{\ell }$ is positive and
\begin{equation}
1\equiv \sum_{\ell }\dim M_{\ell }\mathop{\rm{tr}}\nolimits T_{\ell
}=\sum_{\ell }\left( 2\ell +1\right) \mathop{\rm{tr}}\nolimits T_{\ell
} 
\text{.}  \label{quella}
\end{equation}
It follows that the operators $T$ associated to the $U$-covariant POVMs $M$
by equation~(\ref{L'altra}) are all the operators of the form
\begin{equation}
T=\sum_{\ell }P_{\ell }\otimes T_{\ell }
\end{equation}
with $T_{\ell }$ positive trace class operators satisfying equation~(\ref{quella}).

\section{Proof of theorem \ref{numero}}\label{proof}

We prove theorem \ref{numero} in two steps. First, given an
arbitrary topological group $G$ and a compact subgroup $H$, we
characterize all the POVMs based on the quotient space $G/H$ and
covariant with respect to an irreducible (ordinary) representation
of $G$. Then, we apply the above result to our problem lifting the
projective unitary representation $U$ of the Galilei group to a
(ordinary) unitary representation  of the central extension
$G_\omega$ of the Galilei group defined by the multiplier $\omega$
of $U$.

From now on, let $G$ be a unimodular locally compact second countable topological
group and $H$ be a compact subgroup of $G$. We denote by
\begin{equation*}
G\ni g\longmapsto \pi(g)=\dot{g}\in G/H
\end{equation*}
the canonical projection onto the quotient space $G/H$. Let $\mu
_{G}$ and $\mu _{H}$ be invariant measures on $G$ and $H$
respectively, with $\mu _{H}\left( H\right) =1$. Due to the
compactness of $H$, there exists a $G$-invariant measure $\mu
_{G/H}$ on $G/H$ such that the following measure decomposition
holds
\begin{equation} \int_{G}f\left( g\right) \text{d}\mu
_{G}\left( g\right) =\int_{G/H} \text{d}\mu _{G/H}\left(
\dot{g}\right) \int_{H}f\left( gh\right) \text{d} \mu _{H}\left(
h\right) \text{.}  \label{Mackey-Bruhat}
\end{equation}
for all $f\in L^{1}\left( G,\mu _{G}\right) $.

Let $U$ be an irreducible unitary representation $U$ of $G$ acting
on a Hilbert space $\HH$. We recall that $U$ is said to be
\emph{square-integrable} if there exists a nonzero vector $\phi\in
\mathcal{H}$ such that \begin{equation*} \int_{G}\left|
\left\langle \phi,U_{g} \phi\right\rangle _{\mathcal{H
}}\right| ^{2}\text{d}\mu _{G}\left( g\right) <+\infty \text{.}
\end{equation*}
If the above condition holds, there exists a constant $d_U>0$,
called \emph{formal degree}, such that for all $\phi\in \mathcal{H}$
\begin{equation*}
\int_{G}\left|
\left\langle \phi,U_{g} \phi\right\rangle _{\mathcal{H}}\right| ^{2}\text{d} 
\mu _{G}\left( g\right)
=\frac{1}{d_U} \left\| \phi \right\|^4 \text{.}
\end{equation*}

Finally, all the integrals of operator valued functions (as, for
example, in equation~(\ref{La Povm vecchia}) below) are understood in the weak sense.

We need the following result which is proved in \cite{Nostro}.
\begin{proposition}\label{rivaldo}
Assume that $U$ is  square-integrable with formal degree $d_U$ and let $T$ be positive  trace
class trace one operator $T\in\BB(\HH)$. The map
\begin{equation}
\BB(G)\ni \widetilde{Z}\mapsto \EE_{T}\left( \ZZ\right) =d_{U }\int_{ \widetilde{Z}}U_{g}  T 
U^*_{g}  \text{d}\mu _{G}\left( g\right) \text{,}   \label{La Povm vecchia}
\end{equation}
defines a POVM $\widetilde{E}_T$ on $G$ covariant with respect to $U$.

Conversely, if $\widetilde{E}$ is a POVM on $G$ covariant with respect to $U$, then $U$
is square-integrable and there is  a trace class positive operator
$T\in\BB(\HH)$ with trace one such that $\widetilde{E}=\widetilde{E}_T$.
\end{proposition}

Now we extend the above result to covariant POVMs based on $G/H$.
\begin{corollary}\label{Teo. centr.}
Assume that $U$ is a square-integrable representation with formal degree $d_U$ and let $T$ be a 
trace
class positive operator $T\in\BB(\HH)$ with trace one such that
\begin{equation}\label{commuta}
TU_{h}=U_{h}T\quad\forall h\in H\text{.}
\end{equation}
Then the map
\begin{equation}
\BB(G/H)\ni Z\mapsto E_{T}\left( Z\right) =d_{U }\int_{Z}U_{g} T U^*_{g} 
\text{d}\mu_{G/H}\left(\dot{g}\right) \text{,}   \label{La Povm}
\end{equation}
defines a POVM $E_T$ on $G/H$ covariant with respect to $U$.

Conversely, if $E$ is a POVM on $G/H$ covariant with respect to
$U$, then $U$ is square-integrable and there is  a trace class
positive operator $T\in\BB(\HH)$ with trace one and commuting with
$\left. U \right|_H$ such that $E=E_T$.
\end{corollary}
\begin{proof}
Assume that $U$ is square-integrable and let $T\in\BB(\HH)$ as in the
statement of the corollary. By means of equation~(\ref{La Povm vecchia})
$T$ defines a POVM $\widetilde{E}_T$ based on $G$ and covariant
with respect to $U$. For all $Z\in\BB(G/H)$ let
\[E_T(Z)=\widetilde{E}_T(\pi^{-1}(Z)).\]
Clearly, $E_T$ is a POVM on $G/H$  covariant with respect to $U$.
Moreover, denoting with $\chi_Z$ the characteristic function of
$Z$,
\begin{eqnarray*}
~~~~E_T(Z) & = & d_{U }\int_{G} \chi_Z(\pi(g))U_g T U^*_g  \text{d}\mu _{G}\left( g\right) \\
( \mathrm{eq.~}(\ref{Mackey-Bruhat})) & = &
 d_{U }\int_{G/H} \text{d}\mu _{G/H}\left( \dot{g}\right) \int_{H}
\chi_Z(\pi(gh))U_{gh} T U^*_{gh} \text{d}
\mu _{H}\left( h\right)  \\
( \mathrm{eq.~}(\ref{commuta})) & = &
 d_{U }\int_{G/H} \text{d}\mu _{G/H}\left( \dot{g}\right)
\chi_Z(\dot{g})U_g T U^*_g,
\end{eqnarray*}
that is, equation~(\ref{La Povm}) holds.

Conversely, let $E$ be a POVM on $G/H$ and covariant with respect to
$U$. For all $\ZZ\in\BB(G)$, let $l_\ZZ$ be the function on $G$
given by
\[l_\ZZ(g)=\mu_H(g^{-1}\ZZ\cap H)=\int_H \chi_\ZZ(gh) \text{d}
\mu _{H}\left( h\right).\] Clearly, $l_\ZZ$ is a positive
measurable function bounded by $1$ and, since $\mu_H$ is
invariant, for all $h\in H$, $l_\ZZ(gh)=l_\ZZ(g)$. It follows that
there is a positive measurable bounded function $\ell_\ZZ$ on
$G/H$ such that $l_\ZZ=\ell_\ZZ\circ\pi$.

Define the operator $\EE(\ZZ)$ by means of
\[\EE(\ZZ)= \int_{G/H} \ell_\ZZ(\dot{g}) \text{d}E(\dot{g}),\]
which is well defined since $\ell_\ZZ$ is bounded.

We claim that $\ZZ\mapsto \EE(\ZZ)$ is a POVM on $G$ covariant
with respect to $U$. Clearly, since $\ell_\ZZ$ is positive,
$\EE(\ZZ)$ is a positive operator. Recalling that $\ell_G=1$, one
has $\EE(G)=I$. Let now $(\ZZ_i)$ a disjoint sequence of
$\BB(G)$ and $\ZZ=\cup_i\ZZ_i$. Given $g\in G$, since
$(g^{-1}\ZZ_i\cap H)_i$ is a disjoint sequence of $\BB(H)$ and
$g^{-1}\ZZ\cap H = \cup_i(g^{-1}\ZZ_i\cap H)$, then
$\ell_\ZZ=\sum_i \ell_{\ZZ_i},$ where the series converges
pointwise. Let $\phi\in\HH$, by monotone convergence theorem,
one has that
\[\langle\phi,\EE(\ZZ)\phi\rangle = \sum_i \langle\phi,\EE(\ZZ_i)\phi\rangle.\]
Finally, let $g_1\in G$, then
\begin{eqnarray*}
~~~~\EE(g_1\ZZ) & = & \int_{G/H} \mu_H(g^{-1}g_1\ZZ\cap H) \text{d}E(\dot{g}) \\
( \dot{g}\mapsto g_1\dot{g} ) & = &
\int_{G/H} \mu_H(g^{-1}\ZZ\cap H) U_{g_1}\text{d}E(\dot{g})U^*_{g_1} \\
& = & U_{g_1} \EE(\ZZ) U^*_{g_1}\text{,}
\end{eqnarray*}
where we used the fact that $E$ is covariant.

By means of proposition \ref{rivaldo}, $U$ is square-integrable
and there is a positive trace class operator trace one $T$ such
that
\begin{equation}\label{quasi}
\EE(\ZZ)=d_{U }\int_{\widetilde{Z}}U_{g} T U^*_{g} \text{d}\mu _{G}\left( g\right).
\end{equation}

We now show that $T$ satisfies equation~(\ref{commuta}).  First of all
we claim that, given $h\in H$ and $\ZZ\in\BB(G)$,
\begin{equation}\label{comm_H}
\EE(\ZZ h)=\EE(\ZZ).
\end{equation}
Indeed, since $H$ is compact, $\mu_H$ is both left and right
invariant, so that
\[\mu_H(g^{-1}\ZZ h\cap H)=\mu_H((g^{-1}\ZZ\cap H)h)=\mu_H(g^{-1}\ZZ\cap H)\]
and, hence, $\ell_{\ZZ}=\ell_{\ZZ h}$. By definition of $\EE(\ZZ)$, equation (\ref{comm_H}) easily
follows.
Fixed $h\in H$, by means of equation~(\ref{comm_H}) and equation~(\ref{quasi}) one has that
\begin{eqnarray*}
\int_{ \widetilde{Z}}U_g  T U^*_g
 \text{d}\mu _{G}\left( g\right)
 & = & \int_{ \widetilde{Z}h}U_g  T U^*_g \text{d}\mu _{G}\left( g\right) \\
(\ g\mapsto gh\ )& = & \int_{ \widetilde{Z}}U_{gh}  T U^*_{gh} \text{d}\mu _{G}\left( g\right),
\end{eqnarray*}
where we used the fact that $G$ is unimodular.
Since the equality holds for all $\ZZ\in\BB(G)$, then, for
$\mu_G$-almost all $g\in G$,
\[U_g  T U^*_g=U_g U_h  T U^*_h U^*_g,\] 
where the equality
holds in the weak sense. Since both sides are continuous
functions, the equality holds everywhere and equation~(\ref{commuta})
follows.

Let now $Z\in\BB(G/H)$. Since
\[g^{-1}\pi^{-1}(Z)\cap H=\left\{
\begin{array}{cc}
H & \text{if} \ gH\in Z\\
\emptyset & \text{if} \ gH\not\in Z
\end{array}
 \right. \text{,}\] 
then $\ell_{\pi^{-1}(Z)}=\chi_Z$ and
$\EE(\pi^{-1}(Z))=E(Z)$. Reasoning as in the first part of the
proof one has that $E=E_T$.
\end{proof}

Now we come back to the Galilei group $G$ and to the projective
unitary representation $U$ of $G$ associated with a spinless
particle of mass $m$. We recall that \emph{projective} means that
for all $g_1,g_2 \in G$
\begin{equation*}
U_{g_1}U_{g_2}=\omega(g_1,g_2)U_{g_1 g_2}
\end{equation*}
where $\omega$ is the \emph{multiplier} given by
\begin{equation*}
\omega \left( \left( \vec{a},\vec{v},R\right) ,\left(
\vec{a}^{\prime },\vec{v}^{\prime },R^{\prime }\right)
\right) =e^{im\vec{v}\cdot R\vec{a}^{\prime }}\text{.}
\end{equation*}
We extend $U$ to a unitary representation of the central extension
$G_{\omega }$ of $G$ associated with the multiplier $\omega $
(see, for example, \cite{Var}). Let $\mathbb{T 
}=\left\{ z\in \mathbb{C}:\left| z\right| =1\right\} $ be the multiplicative
group of the torus. The group $G_{\omega }$ is the  product $\mathbb{T} 
\times G$ with the composition law
\begin{equation*}
\left( z,\vec{a},\vec{v},R\right) \left( z^{\prime }, 
\vec{a}^{\prime },\vec{v}^{\prime },R^{\prime }\right)
=\left( zz^{\prime }e^{im\vec{v}\cdot R\vec{a}^{\prime
}},\ \vec{a}+R\vec{a}^{\prime },\vec{v}+R 
\vec{v}^{\prime },\ RR^{\prime }\right) \text{.}
\end{equation*}
In particular, $G_{\omega }$ acts transitively on $X$ by means of
\begin{equation}
\widetilde{\alpha}_{\left(z,\vec{a},\vec{v},R\right)} \left( \vec{
q},\vec{p}\right) =\left( \vec{a}+R\vec{q}
,m\vec{v}+R\vec{p}\right) \text{.}  \label{azione di Gomega}
\end{equation}
and the stability subgroup at the origin is the compact subgroup
$H= \mathbb{T}\times SO\left(3\right) $. In particular, $X$ is
isomorphic to $G_\omega/H$ by means of
\begin{equation}\label{identi}
\left( \vec{
    q},\vec{p}\right)\mapsto
    \pi\left(1,
    \vec{q},\frac{\vec{p}}{m},I\right)\text{,}
\end{equation}
where $\pi:G_\omega\longrightarrow G_\omega /H$ is the canonical
projection.

The irreducible projective representation $U$ lifts to an
irreducible unitary representation $\widetilde{U}$ of $G_{\omega
}$ as
\begin{equation*}
\left[ \widetilde{U}_{\left( z,\vec{a},\vec{v},R\right)}
\phi \right] \left( \vec{x}\right) =z^{-1}e^{im\vec{v} 
\cdot \left( \vec{x}-\vec{a}\right) }\phi \left(
R^{-1}\left( \vec{x}-\vec{a}\right) \right) \text{.}
\end{equation*}
where  $\phi\in L^{2}\left( \mathbb{R}^{3},\text{d}\vec{x}\right) $ .

Clearly a POVM $E$ is covariant with respect to $U$ if
and only if $E$ is covariant with respect to $\widetilde{U}$.
The classification of such POVMs is given in corollary
\ref{Teo. centr.}. We only have to check that the
representation $\widetilde{U}$ is square-integrable (compare with \cite{Ali}). Indeed,
an invariant measure of $G_{\omega }$ is
\begin{equation*}
\text{d}\mu _{G_{\omega }}\left( z,\vec{a},\vec{v} 
,R\right) =\frac{m}{\left( 2\pi \right) ^{3}}\text{d}z\text{d} 
\vec{a}\text{d}\vec{v}\text{d}R\text{,}
\end{equation*}
where d$z$ and d$R$ are normalized Haar measures in $\mathbb{T}$ and in $ 
SO\left(3\right) $ respectively. Moreover, if $\phi \in L^{2}\left( \mathbb{ 
R}^{3},\text{d}\vec{x}\right) $, we have
\begin{eqnarray*}
&&\int_{G_{\omega }}\left| \left\langle \phi ,\widetilde{U}_{\left( z, 
\vec{a},\vec{v},R\right)} \phi \right\rangle \right|
^{2}\text{d}\mu _{G_{\omega }}\left( z,\vec{a},\vec{v} 
,R\right) = \\
&&\qquad =\int_{\mathbb{R}^{3}\times \mathbb{P}^{3}\times SO\left(3\right)
\times \mathbb{T}}\left| \overline{z}\left\langle \phi ,\widetilde{U}_{\left( 1, 
\vec{a},\vec{v},R\right)} \phi \right\rangle \right|
^{2}\frac{m\text{d}\vec{a}\text{d}\vec{v}\text{d}R 
\text{d}z}{\left( 2\pi \right) ^{3}} \\
&&\qquad =\int_{\mathbb{R}^{3}\times \mathbb{P}^{3}\times SO\left(3\right)
}\left| \int_{\mathbb{R}^{3}}\phi \left( \vec{x}\right) e^{-im 
\vec{v}\cdot \left( \vec{x}-\vec{a}\right) } 
\overline{\phi \left( R^{-1}\left( \vec{x}-\vec{a} 
\right) \right) }\text{d}\vec{x}\right| ^{2}\frac{m\text{d} 
\vec{a}\text{d}\vec{v}\text{d}R}{\left( 2\pi \right)
^{3}} \\
&&\qquad =\int_{\mathbb{R}^{3}\times SO\left(3\right) }\left[ {\int_{ 
\mathbb{P}^{3}}}\left| \mathcal{F}\left( \phi \left( \cdot \right) \overline{ 
\phi \left( R^{-1}\left( \cdot -\vec{a}\right) \right) }\right)
\left( m\vec{v}\right) \right| ^{2}m\text{d}\vec{v} 
\right] \text{d}\vec{a}\text{d}R \\
&&\qquad =\int_{\mathbb{R}^{3}\times SO\left(3\right) }\left[ {\int_{ 
\mathbb{R}^{3}}}\left| \phi \left( \vec{x}\right) \overline{\phi
\left( R^{-1}\left( \vec{x}-\vec{a}\right) \right) } 
\right| ^{2}\text{d}\vec{x}\right] \text{d}\vec{a} 
\text{d}R=\left\| \phi \right\| ^{4}\text{.}
\end{eqnarray*}
Then, choosing d$\mu _{G_\omega/H}\left( \vec{a},\vec{v} 
\right) =\frac{m}{\left( 2\pi \right) ^{3}}$d$\vec{a}$d$ 
\vec{v}$, one has $d_{\widetilde{U}}=1$, and every $\widetilde{U}$
-covariant POVM based on $G_\omega/H$ has the form
\begin{equation}
E_{T}\left( Z\right) =\frac{m}{\left( 2\pi \right) ^{3}}\int_{Z} \widetilde{U}_{\left( 
1,\vec{a},\vec{v},I\right)} T\widetilde{U}_{\left(
1,\vec{a},\vec{v},I\right)} ^{\ast }\text{d} 
\vec{a}\text{d}\vec{v}  \label{questa}
\end{equation}
for all $Z\in \mathcal{B}\left( G_\omega/H\right) $, where $T$ is a positive trace
one operator commuting with $\left. \widetilde{U}\right| _{\mathbb{T\times } 
SO\left(3\right) }$. Clearly, $T$ commutes with $\left. \widetilde{U} 
\right| _{\mathbb{T\times }SO\left(3\right) }$ if and only if it commutes
with $ \left. U\right| _{SO\left(3\right) }$. Taking into account the
identification between $X$ and $G_\omega/H$ given by equation~(\ref{identi}), the proof of theorem
\ref{numero} is complete.

\begin{remark}
One can prove that the representation $\widetilde{U}$ is
square-integrable by an abstract argument. Indeed, $G_\omega$ is the
semidirect product of the normal abelian closed subgroup
$\mathbb{T}\times \real^3$ and the closed subgroup $
\mathbb{V}^{3}\times^\prime SO(3)$. Moreover, $\widetilde{U}$ is the
representation unitarily induced by $\sigma$ from
$\mathbb{T}\times \real^3\times SO(3)$ to $G_\omega$, where
$\sigma$ is the representation of $\mathbb{T}\times \real^3\times
SO(3)$ acting on $\mathbb C$ as
\[
\sigma_{(z,\vec x,R)}= z^{-1}\text{.}
\] 
The corresponding orbit in the dual group $\widehat{\mathbb{T}\times
  \real^3}=\mathbb{Z}\times \mathbb{P}^3$ is $\mathcal{O}=\{-1\}\times
  \mathbb{P}^3$. Since $\mathcal{O}$ has a strictly positive measure
  (with respect to the Haar measure of $\mathbb{Z}\times \mathbb{P}^3$)
  and $\left. \sigma\right| _{SO(3)}$ is square-integrable, a theorem
  proved in {\rm \cite{square}} assures that $\widetilde{U}$ is square-integrable.
\end{remark}


\section*{References}
\begin{thebibliography}{12}
 \bibitem{Ludwig-Foundations} {Ludwig G 1983}, {\it Foundations of Quantum
     Mechanics} (Berlin: Springer-Verlag) 

\bibitem{Hol}  Holevo A 1982 \emph{Probabilistic and Statistical Aspects of
Quantum Theory} ( Amsterdam: North-Holland)

\bibitem{pippo2} Busch P, Grabowski M and Lahti P 1997 \emph{Operational Quantum
    Physics} (Berlin: Springer-Verlag)
  
\bibitem{HolevoNEW} {Holevo A 2001} {\it Statistical Structure of
    Quantum Theory} (Berlin: Springer-Verlag) 

\bibitem{HolevoJMP} {Holevo A 1996} Covariant quantum Markovian
  evolutions {\it J.~Math.~Phys.}  {\bf 37} {1812-1832} 

\bibitem{art} {Vacchini B 2001} Translation-covariant Markovian master
  equation for a test particle in a quantum fluid {\it J.~Math.~Phys.}
  {\bf 42} 4290-4312 

\bibitem{art1} {Vacchini B 2002}   
   Quantum optical versus quantum Brownian
  motion master equation in terms of covariance and equilibrium
  properties {\it J.~Math.~Phys. } {\bf 43} 5446-5458
   
\bibitem{Ali-xp}  Ali S T and Prugove\v{c}ki E 1977  Systems of
  imprimitivity and representations of quantum mechanics on fuzzy
  phase spaces {\it J. Math. Phys.} {\bf 18} 219-228 
  
\bibitem{Holevo-xp} {Holevo A 1979} Covariant measurements and
  uncertainty relations  {\it Rep.~Math.~Phys.}  {\bf 16} 385-400 

\bibitem{Nostro}  Cassinelli G, De~Vito E and Toigo A 2003  Positive
operator valued measures covariant with respect to an irreducible
representation {\it J. Math. Phys.} {\bf 44} 4768-4775 

\bibitem{Var}  Varadarajan  V S 1985  \emph{Geometry of Quantum Theory}, II ed.,
(Berlin: Springer-Verlag) 

\bibitem{Ali}  Ali S T 1998 A general theorem on square-integrability: Vector
coherent states {\it J. Math. Phys.} {\bf 39} 3954-3964

\bibitem{square}  Aniello P, Cassinelli G, De Vito E, Levrero A 1998
  Square-integrability of induced representations of semidirect
  products  {\it Rev. Math. Phys}.  {\bf 10}  (1998) 301--313.
  
\end{thebibliography}
\end{document}